\documentclass[sn-mathphys]{sn-jnl}

\usepackage{siunitx}

\theoremstyle{thmstyleone}%
\theoremstyle{thmstyletwo}%

\theoremstyle{thmstylethree}%

\unnumbered

\begin{document}

\title{Ultrabright Entanglement Based Quantum Key Distribution over a 404 km Optical Fiber}

\author[1,2,3]{\sur{Shi-Chang Zhuang}}
\equalcont{These authors contributed equally to this work.}

\author[1,2,3]{\sur{Bo Li}}
\equalcont{These authors contributed equally to this work.}

\author[3,4]{\sur{Ming-Yang Zheng}}
\equalcont{These authors contributed equally to this work.}

\author[1,2,3]{\sur{Yi-Xi Zeng}}

\author[1,2,3]{\sur{Hui-Nan Wu}}

\author[1,2,3]{\sur{Guang-Bing Li}}

\author[3,4]{\sur{Quan Yao}}

\author[3,4]{\sur{Xiu-Ping Xie}}

\author[1,2,3]{\sur{Yu-Huai Li}}

\author[1,2,3]{\sur{Hao Qin}}

\author[5]{\sur{Li-Xing You}}

\author[1,2,3]{\sur{Feihu Xu}}

\author[1,2,3]{\sur{Juan Yin}}

\author[1,2,3]{\sur{Yuan Cao}}

\author[1,2,3,4]{\sur{Qiang Zhang}}

\author[1,2,3]{\sur{Cheng-Zhi Peng}}

\author[1,2,3]{\sur{Jian-Wei Pan}}

\affil[1]{Hefei National Research Center for Physical Sciences at the Microscale and School of Physical Sciences, University of Science and Technology of China, Hefei 230026, China}

\affil[2]{Shanghai Research Center for Quantum Sciences and CAS Center for Excellence in Quantum Information and Quantum Physics, University of Science and Technology of China, Shanghai 201315, China}

\affil[3]{Hefei National Laboratory, University of Science and Technology of China, Hefei 230088, China}

\affil[4]{Jinan Institute of Quantum Technology and CAS Center for Excellence in Quantum Information and Quantum Physics, University of Science and Technology of China, Jinan 250101, China}

\affil[5]{Shanghai Key Laboratory of Superconductor Integrated Circuit Technology, Shanghai Institute of Microsystem and Information Technology, Chinese Academy of Sciences, Shanghai 200050, China}

\abstract{
Entangled photons are crucial resources for quantum information processing. Here, we present an ultrabright polarization-entangled photon source based on a periodically poled lithium niobate waveguide designed for practical quantum communication networks. Using a 780 nm pump laser, the source achieves a pair generation rate of \(2.4 \times 10^{10} \ \text{pairs/s/mW}\). Remarkably, the entangled photons are bright enough to be detected by a power meter, reaching a power of 17.9 nW under a pump power of 3.2 mW. 
We demonstrate the practicality of the source by conducting quantum key distribution experiments over long-distance fiber links. 
\textcolor{black}{Wavelength-division multiplexing was employed to enhance the key generation, and nonlocal dispersion compensation was implemented to ensure precise timing coincidence measurements across a broad spectral range.}
By utilizing  nine pairs of wavelength channels, the system achieved the applicable secure key rates of up to 440.80 \text{bits/s} over 200 km with a 62 dB loss and extended the maximum secure key generation distance to 404 km.  These results demonstrate the potential of wavelength-multiplexed polarization-entangled photon sources for high-speed, long-distance quantum communication, positioning them as key components for future large-scale quantum networks.}

\maketitle

As a fundamental concept in quantum mechanics \cite{pan2012multiphoton}, entanglement is pivotal to various applications within quantum information science, including quantum communication \cite{bouwmeester1997experimental, briegel1998quantum, kimble2008quantum, simon2017towards}, quantum computing \cite{lim2005repeat, walther2005experimental}, and quantum metrology \cite{Demkowicz2014Metrology, riedel2010atom}. 
Specifically, in quantum communication, correlations between entangled photons enable the generation of cryptographic keys with theoretically unbreakable security \cite{ekert1991quantum, bennett1992quantum} and ensure robust operation under diverse conditions \cite{ma2007quantum}. 
Additionally, entanglement underpins the development of various quantum network architectures, including trusted node-free networks \cite{joshi2020trusted} and wavelength-multiplexed quantum networks \cite{wengerowsky_entanglement-based_2018}.

Entanglement-based quantum key distribution (QKD) has achieved significant milestones in both free-space and fiber-based experiments. Low-Earth-orbit satellites extend the distance to over 1000 km \cite{yin2017satellite, Yin2017, yin2020entanglement, Lu2022Micius}. On the ground, optical fibers offer stable and continuous operation unaffected by weather, serving as a flexible complement for serving metropolitan networks. Yet, entanglement-based QKD in fiber is limited by attenuation, resulting in restricted communication distances. Ongoing efforts aim to overcome the limitation \cite{neumann2022continuous, inagaki2013entanglement, wengerowsky_entanglement-based_2018, kimble2008quantum}, with a recent experiment demonstrating the feasibility of QKD over a 248 km fiber link \cite{neumann2022continuous}.
In the future, entanglement-based communication is pivotal for exploring fundamental quantum physics in relativistic regimes in deep space \cite{mohageg2022deep} and for establishing a global quantum network that integrates space and ground components \cite{chen2021integrated, de2023satellite}. Despite these advancements, practical implementations still face challenges. Long-distance demonstrations \cite{yin2020entanglement, neumann2022continuous} remain at the proof-of-principle stage with limited-performance entanglement sources, with asymptotic secure key rates (SKRs) around 1 bit/s and have yet to achieve higher practical key rates. 
\textcolor{black}{Advancing high-speed, long-distance entanglement-based QKD in fiber links requires addressing two key issues: bright entangled photon source and dispersion compensation in long fiber.}

Entangled photon sources have undergone significant development toward practical implementation over the past decades. Early sources, such as beta-barium borate crystals, had achieved a pair generation rate to the order of \(10^4 \ \text{pairs/s/mW}\) \cite{Kwiat1995BBO, Jennewein2000BBO, Aspelmeyer2003BBO, Peng2005BBO, yao2012eightphoton}. 
A breakthrough came with the introduction of quasi-phase-matching (QPM) techniques in periodically poled nonlinear materials, such as periodically poled potassium titanyl phosphate. The generation rate reached around  \( 10^6-10^7 \ \text{pairs/s/mW}\) \cite{Fedrizzi2007,cao2018bell}.
Recent advancements in waveguide technologies have continued to push the boundaries of photon pair generation rates by improving nonlinear interactions through enhanced mode confinement. Periodically poled lithium niobate (PPLN) waveguides, operating at telecom wavelengths, have demonstrated notable advantages, including high spectral brightness \cite{xue2021ultrabright, Kentaro2020ultra}, compactness \cite{vergyris_fully_2017, xue2021ultrabright, zhao2020high} and seamless integration with on-chip technologies \cite{sansoni2017two, Javid2021Chip, atzeni2018integrated, ma2020ultrabright}. \textcolor{black}{Currently, the commonly used waveguides are mainly reverse proton exchange (RPE) waveguides and ridge waveguides. Compared to ridge waveguides, RPE waveguides offer lower losses (less than 0.1 dB/cm), longer lengths, and higher normalized conversion efficiency \cite{Kores2021waveguide}. This work aims to develop a new polarization-entangled photon source based on long RPE waveguides.}

\textcolor{black}{Chromatic dispersion poses a significant challenge in long-distance entanglement distribution. Ultra-low-loss fibers (ULLFs) enable long-range quantum communication \cite{Yin2016MDI, liu2023TF}, and pioneering studies have demonstrated nonlocal dispersion compensation with narrowband sources over a 248 km fiber link \cite{neumann2022experimental}. However, the efficacy of such compensation across a broad wavelength range remains an unresolved experimental challenge. Our research aims to integrate existing dispersion compensation techniques with broadband entanglement in order to mitigate the substantial chromatic dispersion effects.
}

Here, this Letter presents an ultrabright polarization-entangled photon source using a type-0 QPM PPLN waveguide optimized for the practical quantum communication network. 
By employing a 780 nm pump laser, the source can generate nondegenerate parametric photons with a total spectral bandwidth exceeding 60 nm around 1560 nm, allocating 30 nm to signal and idler photons.
With an 18.9 $\mu$W pump power, we derived a pair generation rate of \(4.6 \times 10^{8}\) pairs/s. This result indicates a high pair generation rate per pump power unit of approximately \(2.4 \times 10^{10}\) pairs/s/mW. With a 3.2 mW pump power, we directly measured the parametric photon pairs using a power meter, obtaining a result of 17.9 nW. \textcolor{black}{Our polarization entanglement source has achieved performance levels that significantly surpass the current capabilities of time-resolved detectors for entangled photons. From a practical standpoint, further implementation of dense-wavelength-division multiplexing (DWDM) technology is necessary, or alternatively, advancements in detection technology will be required in the future.}

\textcolor{black}{To demonstrate the source's practical application in quantum communication, we conducted a QKD experiment over long-distance ULLF links. In this experiment, we employed nine pairs of 200 GHz DWDM channels, which spanned approximately 14 nm, to enhance key rates. Within the broad spectral range, the nonlocal compensation of chromatic dispersion issues was effectively addressed.}
Finally,  we achieved SKRs of 440.80 bits/s, 1.87 bits/s, and $1.55 \times 10^{-3}$ bits/s over 201, 301, and 404 km fiber links, respectively, with total losses of 62, 84, and 110 dB. \textcolor{black}{The SKRs are aggregated across all channels. Our results pave the way for entanglement-based long-distance QKD networks.}

\textcolor{black}{Furthermore, the physical impact of the result goes beyond this.}
\textcolor{black}{Future quantum communication networks require intercity nodes, and this work achieves a key rate in the kHz range over fiber links spanning hundreds of kilometers. This brings intercity quantum communication to a truly practical level, reducing the number of communication nodes and resource consumption. Future global QKD networks integrating ground and satellite systems will also rely on high key rates on the ground, a need our entanglement source can effectively meet.} \textcolor{black}{The high-brightness entanglement source demonstrates exceptional attenuation tolerance, reaching up to 110 dB, as verified along fiber links in this study. This remarkable tolerance enables satellite-based quantum experiments over extended distances, such as GEO satellite-to-ground links, Earth-Moon distances \cite{cao2018bell}, and even deep-space scenarios \cite{mohageg2022deep}. These large-scale implementations provide a valuable platform for investigating quantum effects in curved spacetime, contributing to the exploration of the interplay between quantum mechanics and gravity \cite{Rideout_2012}.} 
\textcolor{black}{For the application in quantum information science, high-brightness entangled photon sources boost quantum-enhanced technology development, such as quantum imaging \cite{defienne2024advances}, quantum radar, quantum illumination \cite{Karsa2024radar}, and multiphoton entanglement application \cite{pan2012multiphoton}, by enhancing their performance and expanding their practical applications.}

\begin{figure*}[htbp]   
    \centering
    \includegraphics[width = 12.4cm]{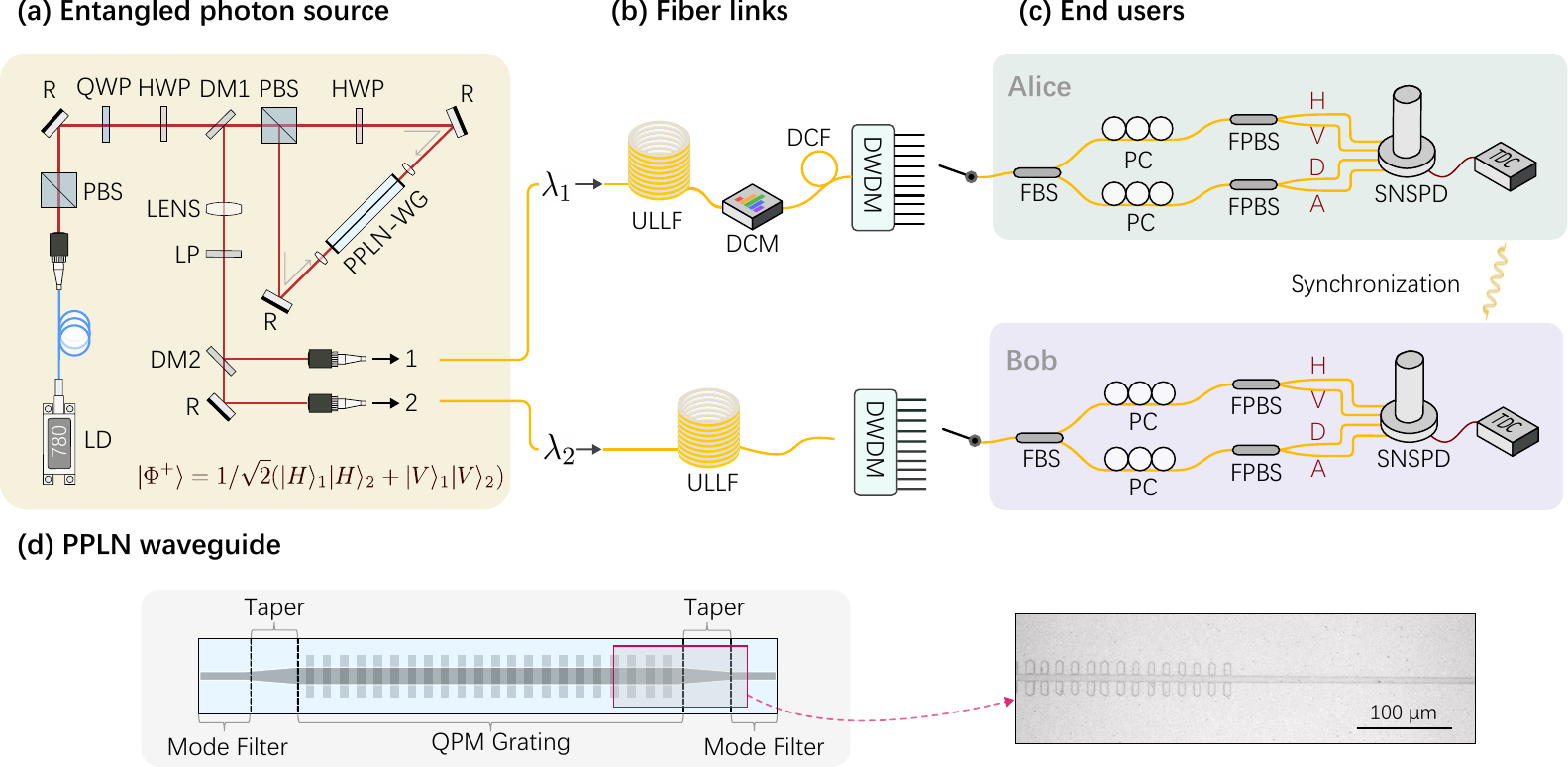}
    \caption{
    Polarization-entanglement source and QKD demonstration.
    (a) Entangled photon source. A CW laser with a wavelength of 780 nm is used to pump the PPLN waveguide in a typical Sagnac ring configuration. The polarization-entangled photon pairs are subsequently collected through single-mode fibers.
    (b) Fiber links. Photon 1 passes through ultra-low-loss fiber (ULLF) and traverses dispersion compensation modules (DCMs) and dispersion compensation fibers (DCFs), which collectively compensate for the chromatic dispersion over the full fiber link. Photon 2 is transmitted directly to the polarization analysis module (PAM) after passing through ULLF. The total length of the two segments of ULLFs is set as 201, 301, and 404 km.
    (c) End users. The nondegenerate entangled photon pairs are split using commercial 200 GHz dense-wavelength-division multiplexing (DWDM) devices that are used separately. For polarization analysis, the photons from each channel are subjected to measurements in randomly chosen $\sigma_z$ and $\sigma_x$ bases.
    (d) The schematic diagram and microscopic image of the PPLN waveguide. The type-0 QPM waveguide is fabricated with the RPE process and optimized for photon generation and coupling.
    LD, laser diode; PBS, polarizing beam splitter; R, reflective mirror; Q, quarter-wave plate; H, half-wave plate; DM1, 780/1560 nm dichroic mirror; PPLN-WG, PPLN waveguide; LP, long-pass filter; DM2, 1540/1580 nm dichroic mirror; FBS, fiber beam splitter; FPBS, fiber polarizing beam splitter; PC, polarization controller; SNSPD, superconducting nanowire single-photon detectors; TDC, time-digital converter.
    The Sagnac interferometer's PBS, H, R, and lens are designed to work at 780 nm and 1560 nm.
    }
    \label{fig:setup}
\end{figure*}

\textcolor{black}{We provide a detailed description of the polarization-entanglement source and QKD system.} The source is shown in Fig.~\ref{fig:setup}(a). It is built using hybrid bulk and waveguide optics techniques in a typical Sagnac configuration \cite{meyer-scott2018highperformance}. This setup includes free-space beam coupling into the waveguide and subsequent coupling of parametric photons into optical fibers.
A continuous-wave (CW) laser with a central wavelength of $\sim$780 nm and a bandwidth of less than 100 kHz serves as the pump source. The laser beam is polarized using a polarizing beam splitter (PBS). Phase compensation in the Sagnac ring is achieved with a quarter-wave plate and a half-wave plate. The pump beam enters the Sagnac interferometer bidirectionally through a PBS, with a half-wave plate in the ring rotating the polarization by 90°. Free-space coupling of the beam to the two input ports of the PPLN waveguide is accomplished using a lens with a focal length of approximately 4.5 mm, with a coupling efficiency of over 50\%. Dichroic mirror 1 (DM1) allows transmission of the pump laser while reflecting the parametric photons. Given the optimal positioning of the coupling lens in the waveguide for focusing the pump beam, an additional lens is placed after DM1 to collimate the parametric photons.
With the waveguide, temperature-controlled, nondegenerate parametric photons are generated and wavelength-tunable (see Appendix or \cite{zhuang2024supp} for details). A long-pass filter blocks the residual pump laser component, and dichroic mirror 2 (DM2) spatially separates the non-degenerate photons according to the spectrum. Finally, the entangled photon pairs are coupled into two single-mode fibers for delivery. 

\textcolor{black}{Specifically, the PPLN waveguide is homemade.} The schematic diagram and microscopic image of the waveguide are shown in Fig.~\ref{fig:setup}(d). \textcolor{black}{In this experiment, the PPLN waveguide is fabricated from z-cut congruent lithium niobate using the RPE process \cite{Parameswaran2002}.}
\textcolor{black}{As the entangled photon pair generation rate ($R$) scales with waveguide length $L$ following the relation $R \propto L^{\frac{3}{2}}$, the waveguide up to 52 mm in length has been developed to enhance the probability of photon pair generation (see Appendix for details). Furthermore, RPE waveguides support single-polarization operation, ideal for type-0 spontaneous parametric down-conversion (SPDC), described by the transition $\vert V\rangle_p \rightarrow \vert V\rangle_1\vert V\rangle_2$,
where $\vert V\rangle$ represents vertical polarization and $p, 1, 2$ denote the pump photon and the generated photon pairs, respectively. This facilitates the utilization of the strongest nonlinear coefficient, $d_{33}$, and enables broadband spectrum generation spanning tens of nanometers \cite{vergyris_fully_2017, xue2021ultrabright}, providing flexibility for multiple wavelength channels from a single source. Such flexibility enhances key generation efficiency \cite{pseiner2021experimental, neumann2022experimental} and supports multiuser quantum network protocols \cite{wengerowsky_entanglement-based_2018}, benefiting quantum networks.}

\textcolor{black}{Waveguide is carefully selected based on its spectral shape, loss characteristics, and conversion efficiency to ensure optimal performance. Additionally, both ends are optimized for free-space coupling to maximize efficiency.}
The waveguide component, fabricated from a 0.5-mm-thick wafer, has a width of 5.5 mm and integrates a mode filter, a taper, as well as a straight waveguide with QPM gratings.
A 1-mm-long 2.5 $\mu$m wide mode filter is located at the input port of the waveguide, followed by a 1-mm-long linear taper, with the waveguide width increasing from 2.5 $\mu$m to 8.0 $\mu$m. 
The QPM grating length is 48 mm, with a poling period of 16.4 $\mu$m. The output end of the waveguide mirrors the input structure, featuring identical mode filters and tapers. Both input and output facets of the PPLN waveguide are antireflection-coated to minimize Fresnel reflection losses.

The generated polarization-entanglement state is close to
$
    \vert \Phi^{+}\rangle=\frac{1}{\sqrt{2}}(\vert H\rangle_1\vert H\rangle_2 + \vert V\rangle_1\vert V\rangle_2),
$
where $\vert H\rangle$ and $\vert V\rangle$ denote horizontal and vertical polarization, and the subscripts 1 and 2 denote the pair of nondegenerate photons, respectively.

\begin{figure}[htbp]   
    \centering
    \includegraphics[width = 8.6cm]{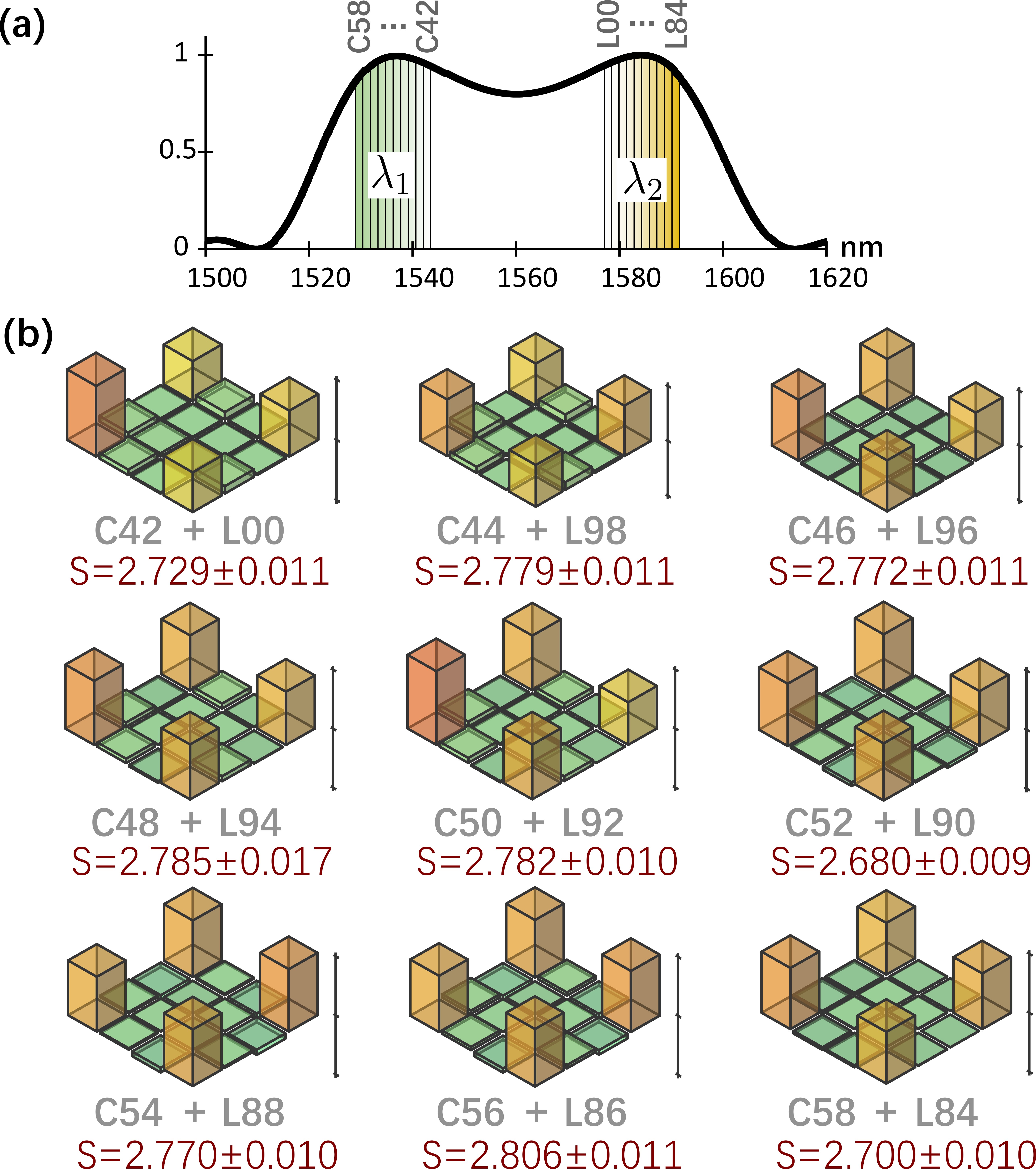}
    \caption{Characteristics of the polarization-entangled photon source.
(a) Spectral characteristics of the source. Through the SPDC process, the entangled photon pairs exhibit a spectral width exceeding 60 nm. In this demonstration, nine pairs of 200 GHz DWDM are selected, ranging from C42 + L00 to C58 + L84, where the channel names denote the center frequency according to the International Telecommunication Union standard. 
(b) Quantum state tomography and S value of Clauser-Horne-Shimony-Holt test for each channel. 
}
    \label{fig:f3}
\end{figure}

\textcolor{black}{The spectrum of the generated photon pairs is shown in Fig. \ref{fig:f3}(a). 
Signal and idler photons show spectral peaks around 1540 nm and 1580 nm, respectively. the parametric photons can cover a bandwidth exceeding 60 nm at specific temperatures, with about 30 nm for each photon, corresponding to a frequency range of over 3.6 THz (see Appendix for details).  
The spectrum of the entangled photons can be divided into DWDMs with a variety of channel spacings. For example, a 200 GHz grid supports approximately 18 pairs of channels, while a 100 GHz grid supports approximately 36 pairs. Narrower channel spacing can increase the number of supported channels.
However, with the increasing number of channels, there is a growing demand for advanced multipixel detection technologies that can efficiently handle simultaneous detections across multiple channels \cite{li2023highrate}.}

The generation rate \( G \), representing the photon pairs generated from the waveguide, is calculated using the formula \( G = \frac{R_c}{\eta^2} \), where \( R_c \) is the detected raw two-photon coincidence count rate \cite{Anwar2021} and \( \eta \) is the average coincidence efficiency. Theoretically, \( G \) is proportional to the pump power \( P \). As depicted in Fig.~\ref{fig:setup}(a), the pump power was measured before the focused beam entered the waveguide.
With a pump power of \( P = 18.9 \, \mu\text{W} \) and an optical attenuator with approximately 2\% transmission, the photon pair coincidence rate was measured as \( R_c = 2.0 \times 10^3 \) pairs/s. When the attenuator is removed, the expected raw pair coincidence rate increases to \( R_c = 5.1 \times 10^6 \) pairs/s with an efficiency of \( \eta = 10.5\% \). Consequently, the generation rate per pump power unit is \( \frac{G}{P} \approx 2.4 \times 10^{10} \) pairs/s/mW, which suggests that a 10 mW pump can generate \( 2.4 \times 10^{11} \) pairs/s. Given that each 1560 nm photon carries a power of \( 1.27 \times 10^{-19} \) W, the emitted photons can reach power levels in the tens of nW range. We directly measured the power of the generated parametric photon pairs after the waveguide using an integrating sphere photodiode power meter. At a pump power of \( P = 3.2 \) mW, the measured power was 17.9 nW, which is close to the expected value of 19.2 nW. This confirms the generation of entangled photons at the tens-of-nanowatt level.

This work used nine pairs of 200 GHz DWDM channels to test the wavelength-multiplexed entanglement characteristics (see Appendix for details). The polarization fidelity, evaluated on two mutually unbiased bases, exceeds 0.99 in average. Figure.~\ref{fig:f3}(b) shows the quantum state tomography \cite{altepeter2005photonic} and the measurements of \( S \) values for  Clauser-Horne-Shimony-Holt type Bell inequalities across the nine channel pairs, demonstrating polarization entanglement properties. The \( S \) value is defined as
$
S = \left\vert E(\varphi_1, \varphi_2) - E(\varphi_1, \varphi_2') + E(\varphi_1', \varphi_2) + E(\varphi_1', \varphi_2') \right\vert,    
$
where \( E(\varphi_1, \varphi_2) \) denotes the correlation function, and \( \varphi_1, \varphi_1' \) (\( \varphi_2, \varphi_2' \)) represent the measurement settings of the photon pairs. The measurements were conducted using the settings \( (0, \frac{\pi}{8}) \), \( (0, \frac{3\pi}{8}) \), \( (\frac{\pi}{4}, \frac{\pi}{8}) \), and \( (\frac{\pi}{4}, \frac{3\pi}{8}) \). The average S value in the Bell test is $2.756\pm0.011$.

In the QKD experiment, entangled photons traveled through two independent ULLF fiber links, as shown in Fig. \ref{fig:setup}(b). \textcolor{black}{Fiber lengths of 201, 301, and 404 km were tested to evaluate performance across varying distances. The ULLF spools are divided into 50 km segments, with attenuation as low as 0.165 dB/km.}
The entangled photons are routed through DWDM channels to reach the users after the fiber links. 

\textcolor{black}{ The dispersion of the ULLF link can be expressed by the following formula:
\begin{equation}
    \sigma_{D}(\lambda) = L_{\mathrm{link}}[D_0 + S_0(\lambda - \lambda_0)],
\end{equation}}
\textcolor{black}{where \(\lambda\) represents the wavelength, \(D_0\) and \(S_0\) are the chromatic dispersion coefficient and chromatic dispersion slope coefficient at the reference wavelength \(\lambda_0\), and \(L_\mathrm{link}\) denotes the fiber link length. In our experiment, \(D_0\) is 17 ps/nm/km, \(S_0\) is 0.06 ps/nm\(^2\)/km, and \(\lambda_0 = 1550\) nm.  
We adopt the chromatic dispersion compensation method proposed by Franson \cite{Franson1992, Catherine2014}, which enables nonlocal addressing of dispersion for entangled photons. In our experiment, we integrated two types of commercial compensation devices: dispersion compensation modules (DCMs) and auxiliary dispersion compensation fibers (DCFs). The DCMs, based on fiber Bragg gratings, are particularly effective for mitigating dispersion over hundreds of kilometers, while the DCFs are ideal for compensating residual dispersion over tens of kilometers. By adding these two types of devices only on the Alice side, we achieve precise control of dispersion compensation.   
}

\textcolor{black}{
The effectiveness of chromatic dispersion compensation in our experiment is demonstrated through two-photon time correlation measurements, which reflect the system's total timing uncertainty \cite{Neumann2021Nonlocal,neumann2022experimental}. The total timing uncertainty can be expressed as:
\begin{equation}
    \Delta T(\lambda_1,\lambda_2) = \sqrt{\sigma_0^2 + \left(\sigma_D^A(\lambda_1) + \sigma_D^B(\lambda_2) + \sigma_C(\lambda_1)\right)^2},
\end{equation}}
\textcolor{black}{
where \(\sigma_0\) represents the initial timing uncertainty in local coincidence measurements, arising from the jitter of the superconducting nanowire single-photon detectors (SNSPDs), time-digital converters (TDCs), and the clock board, which is measured to be 60 ps. \(\sigma_D^A(\lambda_1)\) and \(\sigma_D^B(\lambda_2)\) denote the dispersion contributions from the respective fiber links, while \(\sigma_C(\lambda_1)\) corresponds to the applied dispersion compensation at the 1540 nm end.
}

\textcolor{black}{
For fiber links of 201, 301, and 404 km across different DWDM channels, \(\sigma_C(\lambda_1)\) ranges from \(-4526\) to \(-4324\) ps, \(-6580\) to \(-6287\) ps, and \(-9058\) to \(-8654\) ps, respectively. 
Typically, for a 301 km link, dispersion compensation is optimized in the central DWDM channel. Consequently, in this channel \(\Delta T \approx 80\) ps and spans 80 to 130 ps across the full spectrum due to compensation imperfections and higher-order dispersion. Details of the configurations as well as the setups for Alice and Bob, are provided in Appendix. Improved performance requires further precise compensation for each channel, along with customized, loss-optimized devices.
}

At the end users in Fig.~\ref{fig:setup}(c), passive polarization analysis modules consisting of 50:50 beam splitters (BSs) and polarizing beam splitters (PBSs) are employed to generate the key. The photons are measured in two mutually unbiased bases: horizontal/vertical (H/V) and diagonal/antidiagonal (D/A) linear polarization bases. Each user operates an independent detection system, including low-jitter SNSPDs and TDCs, with time synchronization achieved via an atomic clock and a clock board.

QKD data were acquired from different pairs of DWDM channels separately. Utilizing the secure analysis framework from~\cite{lim2021SKR}, we derived the SKR for fiber lengths of 201 and 301 km. 
We selected a photon pair generation rate of approximately $5.5 \times 10^8 $ pairs/s per channel to optimize the key generation rate (see Appendix for details). Accordingly, the pump power is around 0.55 mW. \textcolor{black}{In the below analysis, the SKRs, raw key lengths, and final key lengths are aggregated across all channels.}

For the 201 km fiber link, with an average data acquisition time of 9.8 min per channel, 1 353 038 bits of raw key were collected with an overall loss of 62 dB. The quantum bit error rate (QBER) was 5.12\%. The final secure key consisted of 130,845 bits, yielding an SKR of 222.98 bits/s. For the 301 km fiber link, with an average acquisition time of 81.1 minutes per channel, 65 764 bits of raw key were obtained under an overall loss of 84 dB. The QBER increased to 6.40\%. After secure key analysis, the final secure key contained 2534 bits, resulting in an SKR of 0.52 bits/s.

For the 404 km fiber link, with an average measurement time of 333.3 minutes, we collected 552 bits of raw key under a total loss of 110 dB. \textcolor{black}{The sifted key rate was 0.014 keys per second, and the measured QBER was 8.69\%, primarily influenced by the low count rate due to the high channel loss.
}

We performed an asymptotic key rate analysis (see Appendix for details) and generated a final secure key of 31 bits, corresponding to an asymptotic SKR of \(1.55 \times 10^{-3}\) bits/s. \textcolor{black}{Because of the limited amount of sifted key, finite-key analysis was not conducted in the 404 km fiber link.} 

\begin{figure}[htbp]
    \centering
    \includegraphics[width=10cm]{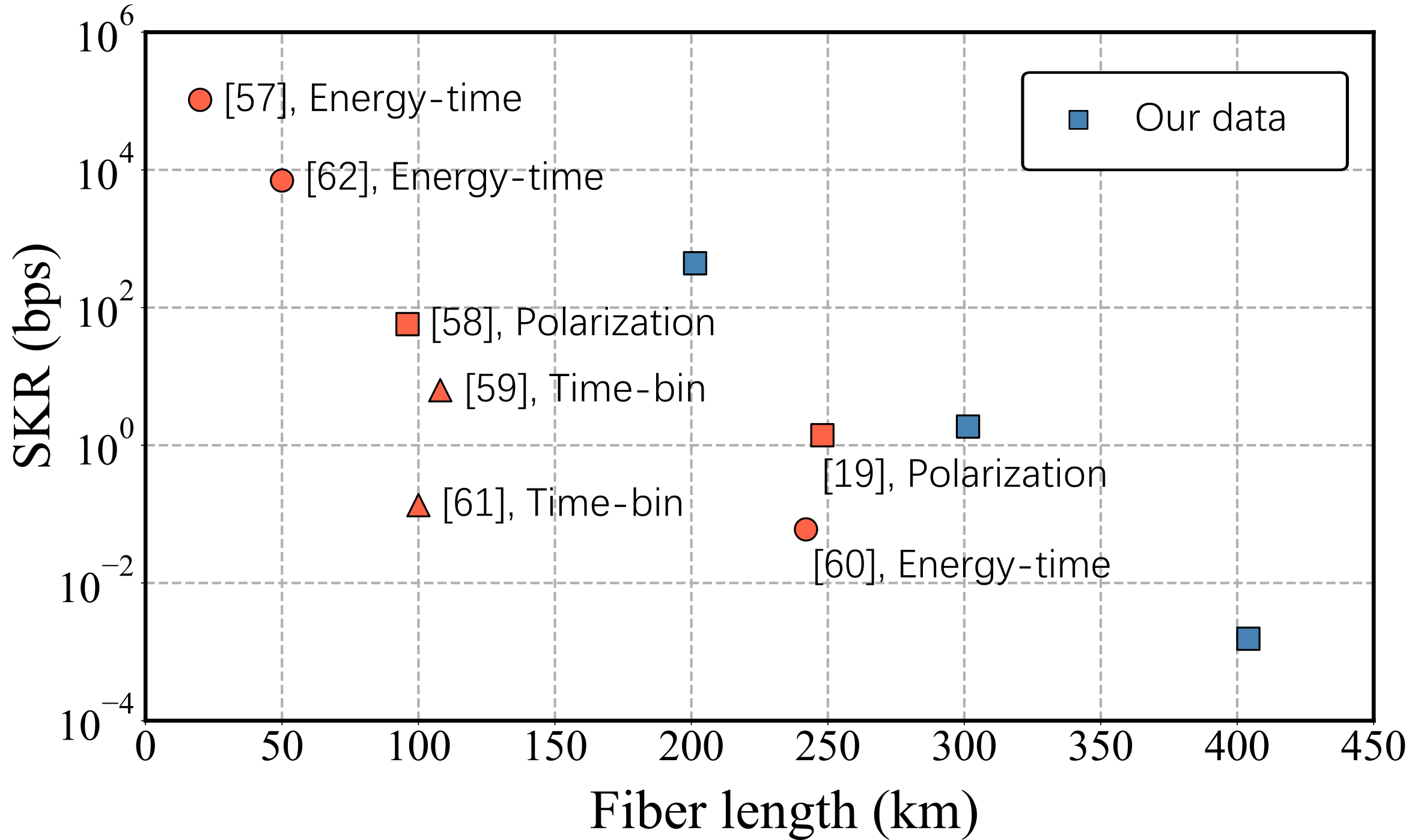}
    \caption{
    Experimental data compared to others. We compared our experimental data with other studies and comprehensively reviewed advancements in fiber-based entanglement QKD implementations. Our key generation results for three distinct distances exhibit an advantage over existing benchmarks~\cite{Liu2019,Wengerowsky2019,Fitzke2022,Liu2024,Honjo08,neumann2022continuous,Pelet2023}. The results of the polarization-entangled QKD experiments were analyzed by 
    asymptotic key rate formula. 
    }
    \label{fig:data-compare}
\end{figure}

Additionally, we calculated the asymptotic key rates for the 201 km and 301 km fiber links.
 For the 201 km fiber link, the asymptotic secure key was 258 663 bits, yielding a key rate of 440.80 bits/s. For the 301 km fiber link, the asymptotic secure key was 9088 bits, resulting in a key rate of 1.87 bits/s. The statistical results are summarized in Table~\ref{tab:QKD}. We compared our data with other recent works in Fig. \ref{fig:data-compare}, highlighting our advantage in SKR. The asymptotic key rate results for distances of 201, 301, and 404 km show an improvement, benefiting from multiplexing. 

\begin{table*}[htbp]
    \centering
    \caption{\bf Key analysis for QKD. }
    \begin{tabular}{crrr}
    \hline
    \hline
    Item & 201 km & 301 km & 404 km\\
    \hline
    Loss (dB) & 62 & 84 & 110\\
    Data acquisition time (min) & 9.8 & 81.1 & 333.3\\
    QBER ($\%$) & 5.12 & 6.40 & 8.69\\
    Raw key (bit) & 1 353 038& 65 764& 552\\
    Sifted key (bit) & 676 519& 32 882& 276\\
    Secure key (bit) & 130 845& 2534& \\
    Secure key rate (bit/s) & 222.98 & 0.52 & \\
    Asymptotic key (bit) & 258 663& 9088& 31\\
    Asymptotic key rate (bit/s) & 440.80 & 1.87 & $1.55 \times 10^{-3}$\\
    \hline
    \hline
    \end{tabular}
    \label{tab:QKD}
\end{table*}

Our work demonstrates significant scalability and is well-positioned for future enhancements, particularly with the use of narrower DWDM bandwidths and multipixel SNSPDs. When deployed in low-Earth-orbit satellite-to-ground QKD systems \cite{yin2020entanglement}, this source can achieve a practical key rate on the order of kbit/s, thus enhancing the feasibility of satellite-based quantum communication, despite a loss range of 60-70 dB \cite{Yin2017,Lu2022Micius}. Moreover, the source's ultrahigh generation rate makes it particularly suitable for future applications in medium- and high-Earth-orbit satellite scenarios, where extra losses of approximately 30 dB are anticipated. Additionally, PPLN waveguides offer notable advantages in both brightness and compactness. High-brightness entanglement sources based on PPLN are progressing toward miniaturization, with future research aiming to verify their performance in satellite applications. Notably, the brightness of the entangled source in this work reaches \(2.4 \times 10^{10} \ \text{pairs/s/mW}\), which already approaches the temporal resolution limit of current single-photon detectors and electronic systems. With future advancements in electronic technologies and further optimization of dispersion compensation techniques, the scope of long-distance QKD experiments can be substantially expanded.

\section*{Acknowledgments}
We acknowledge the experimental support from Ting Zeng, Zhi-Xuan Geng, Si-Yuan Sun, Li-Kang Zhang, and Yang Liu. This work has been supported by the National Natural Science Foundation of China (Grants No. 92476203, No. T2125010, No. 12274398, No. 12174374, No. 62031024), the National Key R\&D Program of China (Grant No. 2020YFA0309803), the Innovation Program for Quantum Science and Technology (Grants No. 2021ZD0300100, No. 2021ZD0300800, No. 2023ZD0300100), the Chinese Academy of Sciences (CAS), Shanghai Municipal Science and Technology Major Project (Grant No. 2019SHZDZX01), CAS Young Inter-disciplinary Innovation Team (Grant No. JCTD-2022-20), the CAS Project for Young Scientists in Basic Research (Grants No. YSBR-046, No. YSBR-085), the Key R\&D Plan of Shandong Province (Grant No. 2021ZDPT01), Natural Science Foundation of Shandong Province (ZR2021LLZ013, ZR2022LLZ009), and Anhui Initiative in Quantum Information Technologies. Y.-H. Li. was supported by the Youth Innovation Promotion Association of CAS (under Grant No. 2023475).


\newpage

\section*{Appendix}

\subsection*{S1. SPDC spectrum in PPLN waveguide}

\begin{figure}[htbp]
    \centering
    \includegraphics[width=10cm]{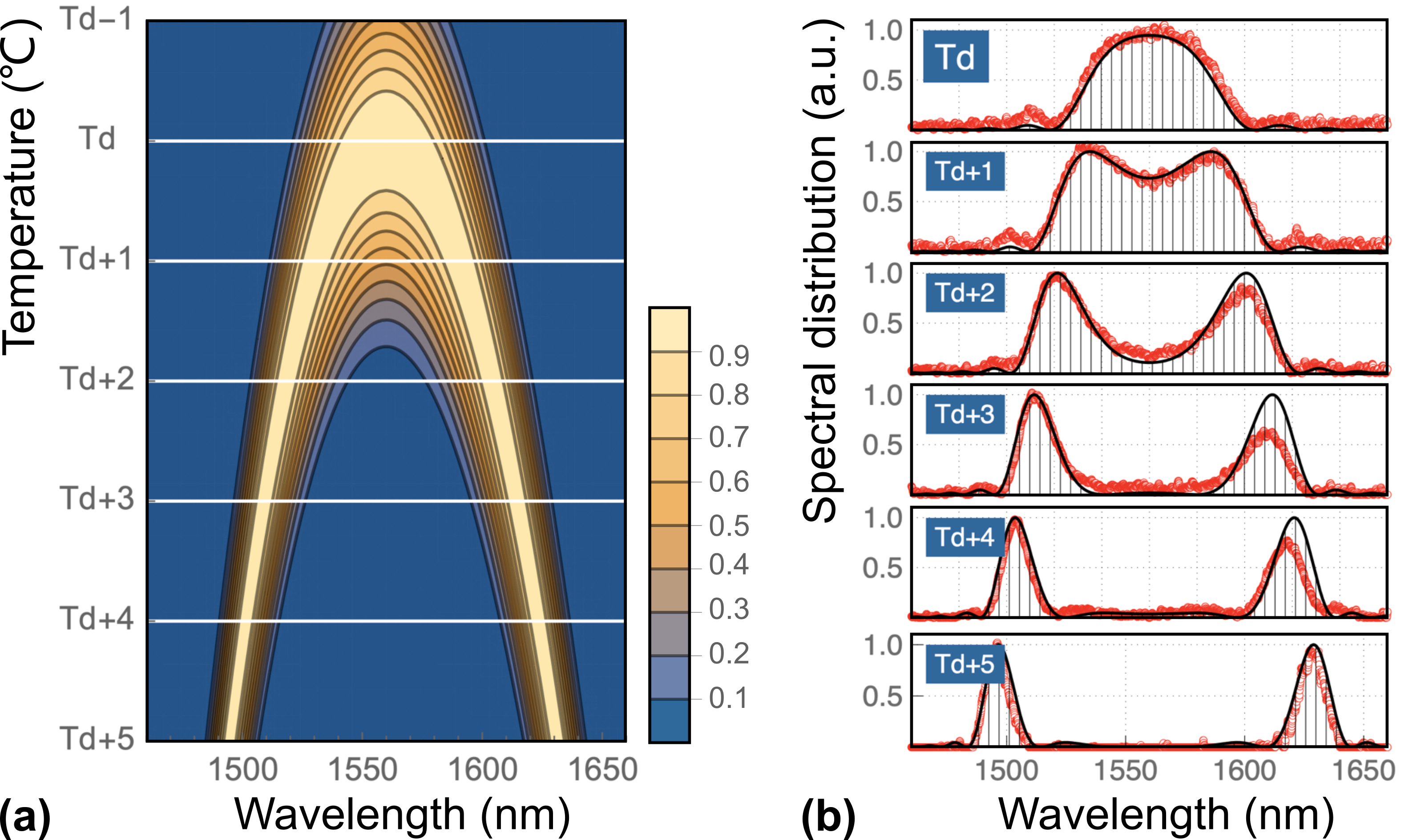}
    \caption{
    Temperature-dependent spectral characterization of photon pairs generated in PPLN waveguide.
    (a) SPDC spectral distribution in theory. 
    The contour plot depicts the wavelength of generated photons as a function of the temperature. 
    The intensity in the graph is a normalized representation with arbitrary units (a.u.). 
    $T_d$ is the working temperature of PPLN at the degenerate point. 
    (b) The measured spectrum. The data (red dots) shows the spectrum trend with temperature, which agrees with the theoretical expectation (solid black line).
    }
    \label{fig:temptuning}
\end{figure}

The down-converted photons and pump photons satisfy energy conservation: 
$1/ \lambda_p=1/\lambda_1+1/\lambda_2$,
where $\lambda_p, \lambda_1,\lambda_2$ denotes the wavelength of the pump and non-degenerate photon pairs, respectively. 
The spectral emission rate can be estimated as
\begin{equation}
    R(\lambda) \propto [\text{sinc}(\frac{L}{2}\Delta k)]^2,
    \label{equ:phase-matching}
\end{equation}
where $L$ is the waveguide length and $\Delta k = 2\pi(\frac{n_p}{\lambda_p} -  \frac{n_1}{\lambda_1} - \frac{n_2}{\lambda_2} - \frac{1}{\Lambda})$. 
The poling period of the periodically poled lithium niobate (PPLN) waveguide, denoted by $\Lambda$, and the refractive index of the lithium niobate material, denoted by $n$, which is a function of wavelength and temperature, are parameters that determine the spectral distribution of parametric photons in PPLN. 
As the operating temperature of the PPLN waveguide increases, the peak of the SPDC photon spectrum shifts from the degenerate wavelength near 1560 nm toward the spectral edges. The total spectrum of the photon pairs spans a range exceeding 60 nm, with the full wavelength at half maximum (FWHM) progressively narrowing as the temperature rises.

The detailed characteristics of the spectral tuning with temperature are depicted in Fig.~\ref{fig:temptuning}(a). We employed an optical spectrum analyzer (OSA) to directly measure the SPDC photon spectrum following the excitation of the PPLN with a 10 mW laser. The OSA had a resolution of 1 nm with a sweep step size of 0.1 nm. Residual pump power was filtered out before measurement with the OSA.

By adjusting the working temperature of the PPLN, the spectrum can be extensively tuned to cover the C+L band and beyond. The SPDC photons are emitted only above a certain degenerate temperature point (\(T_d\)). As the temperature increases, the spectrum of the photon pairs undergoes splitting and shifts toward both spectral edges. The central wavelength changes by approximately 10 nm/°C. As the spectrum shifts further from 1560 nm, the FWHM of the spectrum narrows from roughly over 30 nm at \(T_{d+1}\) to around 10 nm at \(T_{d+5}\).

\begin{figure}[htbp]
    \centering
    \includegraphics[width=10cm]{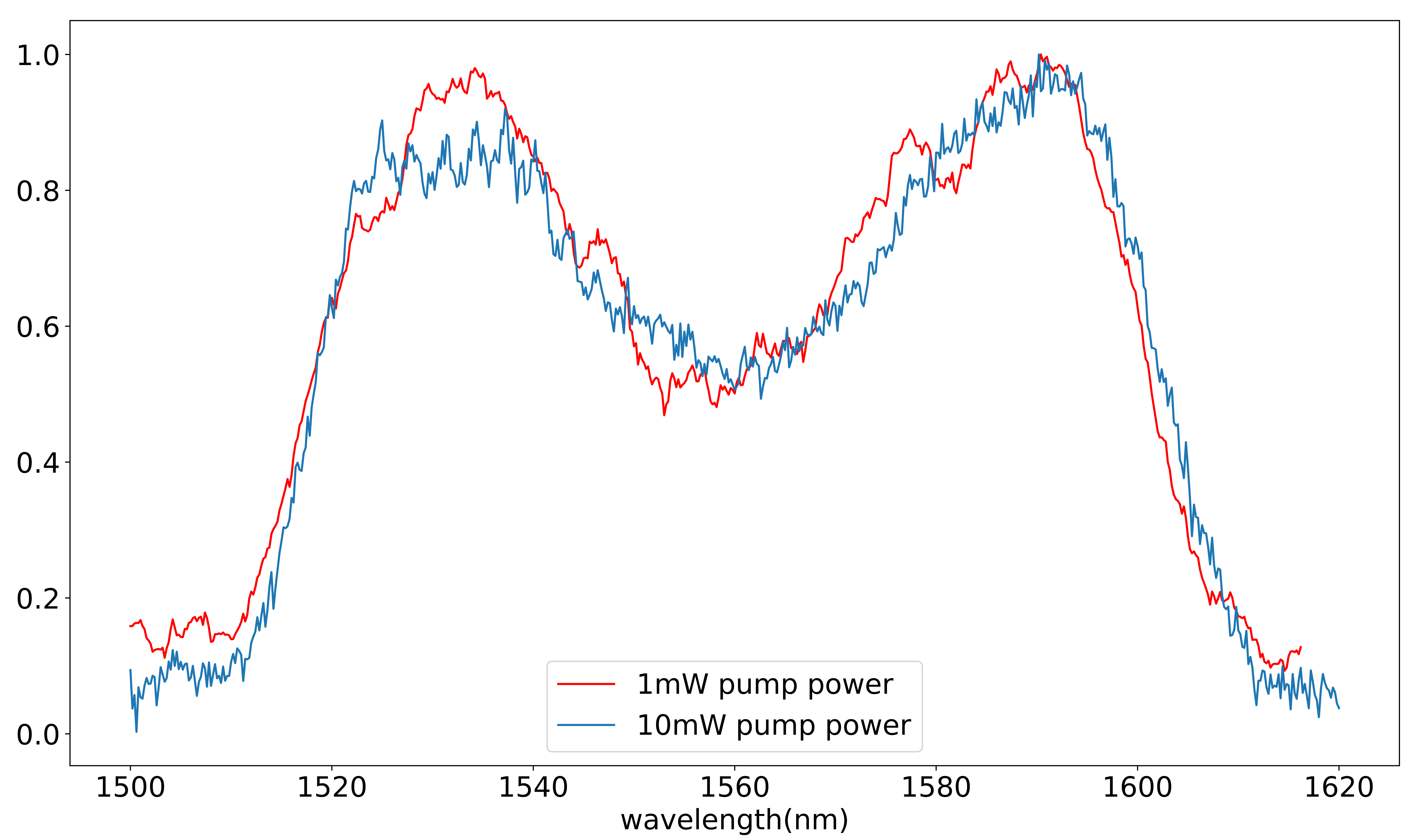}
    \caption{\textcolor{black}{Spectral characterization under different pump powers. The SPDC spectral distribution in \(T_{d+1.5}\) for pump powers of 1 mW and 10 mW exhibits no additional nonlinear effects dependent on the pump power. The vertical axis is normalized.}
    }
    \label{fig:pump}
\end{figure}

\textcolor{black}{Furthermore, the spectra under different pump powers of 1 mW and 10 mW are presented in Fig. \ref{fig:pump}. The spectra show no additional nonlinear effects dependent on the pump power.}

\subsection*{S2. Test on spectral generation rate}

To verify the spectral generation rate, we employed a pair of 200 GHz DWDM filters to select the parametric photon pairs within a FWHM of approximately $1.25 \ \text{nm}$. With a pump power of $P=7.0$ $\mu$$\text{W}$, we observed a two-photon coincidence count rate of $R_c=8.0 \times 10^4 \ \text{pairs/s}$, with $\eta =10.7\%$. 
As expected, the resulting spectral photon generation rate is approximately $8.0 \times 10^8 \ \text{pairs/s/nm/mW}$.

\subsection*{S3. Fabrication process of the PPLN waveguide}

The fabrication process of PPLN waveguides consists of two main stages: the first stage involves the periodic poling of the lithium niobate crystal, while the second stage entails the formation of the optical waveguide structure within the crystal.

Periodically poling involves inducing a periodic variation in the polarization direction of the lithium niobate wafer, alternating between upward and downward polarization, by applying a strong electric field at both ends. We used a 3-inch diameter wafer with a thickness of 0.5 mm. The process consists of the following steps:
1. Spin-coat a thin layer of photoresist (several micrometers thick) on the +Z face of the wafer, followed by prebaking.
2. Define the periodic electrode pattern using a polarization mask and photolithography system, followed by UV exposure, development, and post-baking.
3. After cooling, apply a 10.5 kV electric field to induce polarization reversal.
4. Etch the wafer's surface and edge regions using hydrofluoric acid.

The waveguide structure is realized by introducing a waveguide region with an increased refractive index in the periodically poled lithium niobate crystal, ensuring light confinement. The waveguide fabrication process is as follows:
1. Deposit a silica thin film on the -Z face of the lithium niobate wafer using physical vapor deposition (PVD).
2. Spin-coat a layer of photoresist (several hundred nanometers thick) and prebake in an oven.
3. Define the waveguide pattern using a waveguide mask and photolithography system, followed by UV exposure, development, and post-baking.
4. Etch the exposed silica using buffered oxide etch (BOE) solution to form the waveguide path.
5. Dice the wafer into waveguide chips with dimensions of 52 mm by 5.5 mm.
6. Perform a series of processes, including proton exchange (PE), annealing, Reverse proton exchange (RPE), and polishing, to create the RPE waveguide structure.

PE involves replacing lithium ions with hydrogen ions in a liquid benzoic acid solution, with an exchange time of 18 to 24 hours. The annealing process allows the protons and lithium ions to diffuse, restoring optical nonlinearity and improving mode matching across different wavelengths. This is done by placing the chip in a three-zone tube furnace at approximately 300 $^\circ$C for thermal diffusion, followed by natural cooling. RPE improves the refractive index distribution in the waveguide region for better mode matching. The chip is placed in a molten buffer solution of $\mathrm{LiNO_3}$, $\mathrm{NaNO_3}$, and $\mathrm{KNO_3}$ at around 300 $^\circ$C, enabling $\mathrm{Li^{+}}$ diffusion to optimize the hydrogen ion concentration distribution along the depth of the waveguide.

\subsection*{S4. Challenges and restrictions in PPLN waveguide}

The RPE waveguide serves as the key component for ultra-high brightness through optimized nonlinear interactions. Theoretically, brightness scaling depends on several key parameters, such as cross-sectional area, group velocity dispersion, and waveguide length L. For RPE waveguides, the width of the waveguide is chosen based on the noncritical dimension, and the group velocity dispersion is difficult to engineer because of weak evanescent wave effect, so L dominates due to the dependence of \( P \propto L^{3/2} \). While lengthening L is theoretically optimal, practical implementation faces critical technical limitations: temperature gradients during proton exchange induce inhomogeneous ion diffusion, resulting in refractive index fluctuations that disrupt quasi-phase-matching conditions. These cumulative defects ultimately suppress brightness enhancement despite increased L. For example, a temperature deviation of ± 0.02°C can cause approximately a 1 nm phase-matching wavelength shift, which exceeds the phase-matching bandwidth ($\sim $ 0.3 nm). Our current fabrication process achieves 52-mm-long waveguides on 3-inch wafer. Extending beyond this length requires further improvements in furnace temperature uniformity on larger wafers.

\subsection*{S5. Wavelength-multiplexed entanglement distribution architecture}
\begin{figure}[htbp]   
    \centering
    \includegraphics[width =10cm]{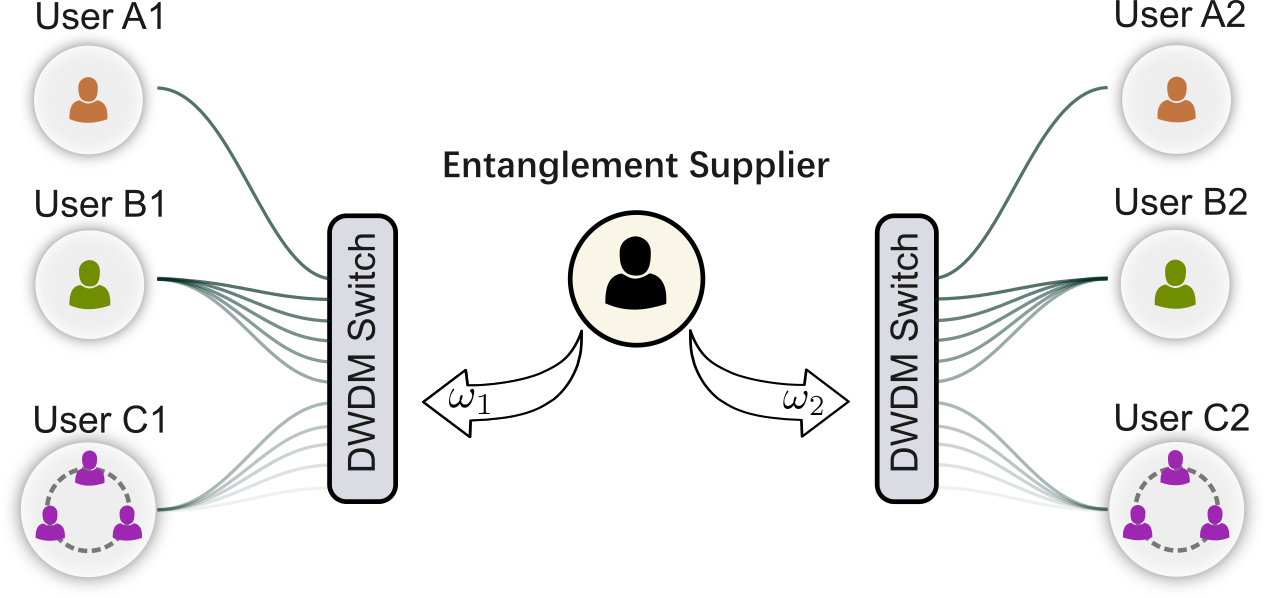}
    \caption{Wavelength-multiplexed quantum entanglement distribution.
    The entanglement supplier possesses an entangled photon source with a broad bandwidth and distributes photons to remote nodes. A DWDM switch is an intermediary router, segmenting the bandwidth into narrow wavelength channels based on user requirements. Users can communicate within a single channel, such as User A1 and A2, or, for high-speed key generation, users can simultaneously utilize multiple channels, such as B1 and B2. Furthermore, multiplexing can maximize the utilization of entanglement resources, enabling the construction of a flexible quantum network that connects various communication groups, such as C1 and C2.
    }
    \label{fig:network}
\end{figure}
The architecture that leverages the broad-spectrum capability of our entangled photon source is illustrated in Fig.~\ref{fig:network}. It facilitates entanglement distribution across multiple wavelengths, rendering it highly suitable for advanced high-speed, secure quantum cryptography and practical multi-user network implementations.

\subsection*{S6. Time jitter analysis}

\textcolor{black}{Total time jitter of coincidences primarily arises from three components: the time jitter of the superconducting single-photon detectors (SNSPDs), time-to-digital converters (TDCs), and clock boards used to synchronize the two TDCs. The TDCs we use are from Swabian Instruments, specifically the performance version of the Time Tagger Ultra, with a single-channel RMS time jitter of 8 ps and an inter-channel FWHM time jitter of approximately 25 ps. The FWHM time jitter for the clock boards and TDCs combined is 25 ps. Two different SNSPDs were used in the experiment with corresponding single-channel time jitters of 30 ps and 45 ps, respectively. The final time jitter for the SNSPDs, clock boards and TDCs combined is 60 ps. We use the same set of dispersion compensation equipment to compensate all DWDM channels, which leads to inconsistent dispersion compensation results for different channels. We optimized the compensation for the middle channel, while the time jitter for the edge channels is broadened.}

\subsection*{S7. Optimization of SKR}
\begin{figure}[htbp]
    \centering
    \includegraphics[width=10cm]{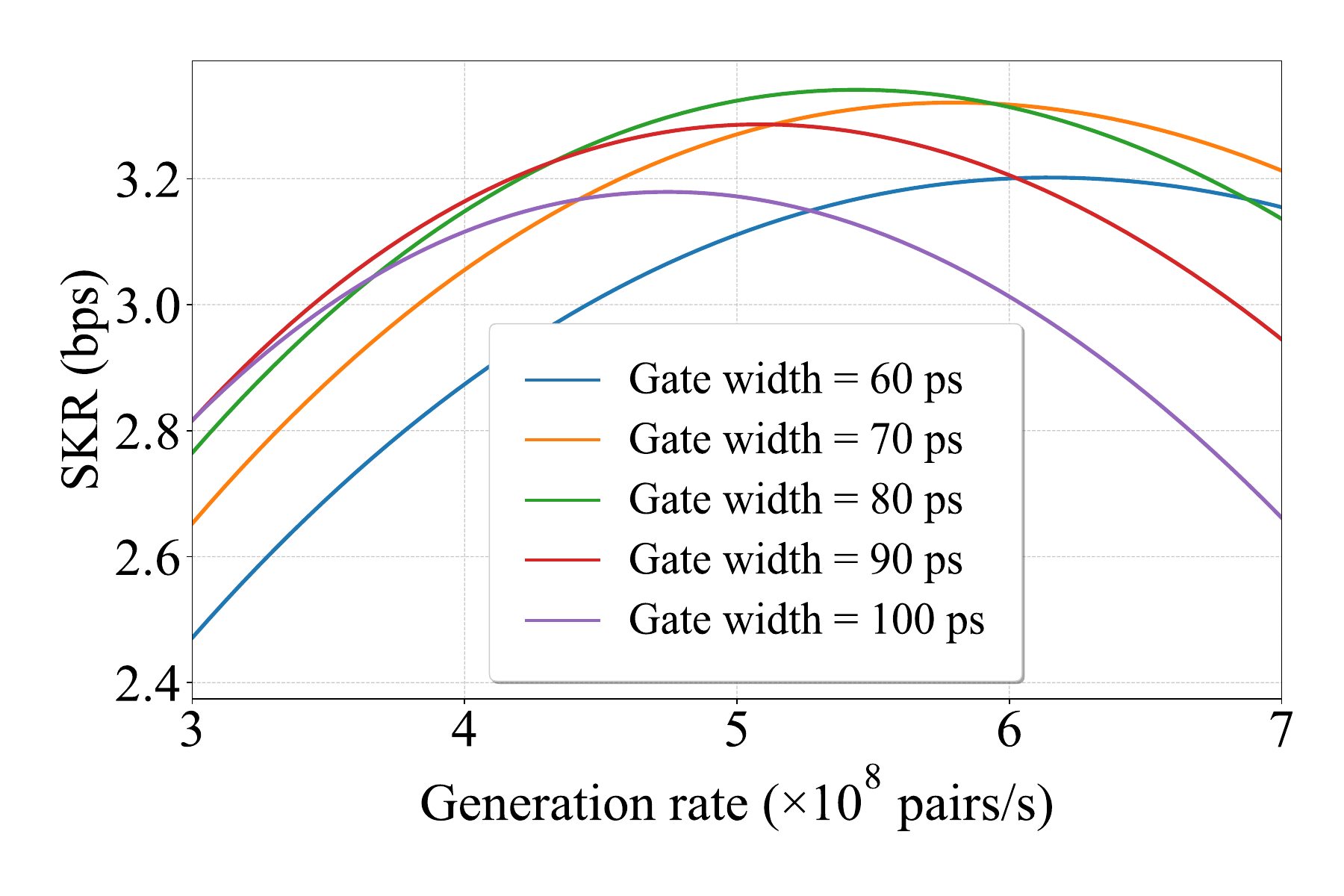}
\caption{
Optimization of the key rate as a function of the generation rate. The error correction leakage factor is set at \(f_e = 1.09\). The analysis uses experimental loss data for a 301 km fiber link, with a coincidence window FWHM of 65 ps as an example to estimate the SKR. The colored curves represent SKR values for various coincidence window gate widths applied to the same dataset. The optimal key rate is observed at a generation rate of \(5.5 \times 10^8\) pairs/s with a gate width of 80 ps.
}
    \label{fig:gen}
\end{figure}

Increasing pump power can enhance the generation rate of entangled photon pairs, however, the achievable secure key rate (SKR) for each channel is ultimately limited by the system's time jitter, which affects the quantum bit error rate (QBER) due to accidental coincidences. To identify the optimal generation rate per channel, we conducted numerical simulations.
We determined the generation rate that maximizes the SKR by applying the asymptotic key length formula. The simulation results, presented in Fig.~\ref{fig:gen}, indicate that the optimal single-channel generation rate is \(5.5 \times 10^8\) pairs/s.

\subsection*{S8. Dispersion compensation}

\textcolor{black}{Two main dispersion effects influence the experiment: chromatic dispersion (CD) and polarization mode dispersion (PMD). CD is proportional to the fiber length \( L \) and the signal's spectral width, and it is the dominant dispersion effect in fiber link. In addition, the PMD is proportional to \( \sqrt{\text{L}} \), resulting in a value of less than 0.04 ps/\( \sqrt{\text{km}} \). Even over a distance of 400 km, the PMD remains below 0.8 ps, which is significantly smaller than the estimated photon coherence time of approximately 6 ps. This value is negligibly small, and its impact on the experimental results can thus be considered insignificant. Therefore, in the following discussion, we will focus on the impact of CD.}

When entangled photon pairs traverse a dispersive medium, the time correlation function \(\sigma_{cor}\) within the interference time becomes \(\sigma'_{cor}\) :

\textcolor{black}{
\begin{equation}
    \sigma{'}_{cor}^2 \approx \frac{1}{\sigma_{cor}^2}[\sigma_{cor}^4+(\beta_A L_A + \beta_B L_B)^2],
\end{equation}}

where \(\beta_{\{A,B\}}\) represents the group-velocity dispersion (GVD) introduced by Alice and Bob over the fiber lengths \(L_{\{A,B\}}\).

\textcolor{black}{
According to the ITU standard, the dispersion of ultra-low-loss fiber (ULLF) link can be expressed using the following formula:}

\textcolor{black}{
\begin{equation}
    \sigma_{link}(\lambda) = L_{link}[D_0+S_0(\lambda - \lambda_0)].
\end{equation}}

\textcolor{black}{The $\lambda$ represents the wavelength. $D_0$ and $S_0$ are the chromatic dispersion coefficient and chromatic dispersion slope coefficient, respectively, at the wavelength of $\lambda_0$. The fiber link length is $L_{link}$.}

\textcolor{black}{We used DCMs and DCFs to compensate for the total dispersion.
All DCMs and DCFs were placed at the 1540 nm side. Correspondingly, the dispersion compensation coefficients and dispersion compensation slope coefficients of DCF and DCM as a function of wavelength can also be expressed using a similar formula.}

\textcolor{black}{In the 201 km QKD experiment, we placed 100 km of ULLF at the 1540 nm side, using one DCF and two DCMs for dispersion compensation, while 100 km of ULLF was placed at the 1580 nm side. In the 301 km QKD experiment, we placed 200 km of ULLF at the 1540 nm side, using one DCF and three DCMs for dispersion compensation, while 100 km of ULLF was placed at the 1580 nm side. In the 404 km QKD experiment, we placed 200 km of ULLF at the 1540 nm side, using three DCFs, four DCMs for dispersion compensation, while 200 km of ULLF was placed at the 1580 nm side.}

\begin{figure}[htbp]   
    \centering
    \includegraphics[width=10cm]{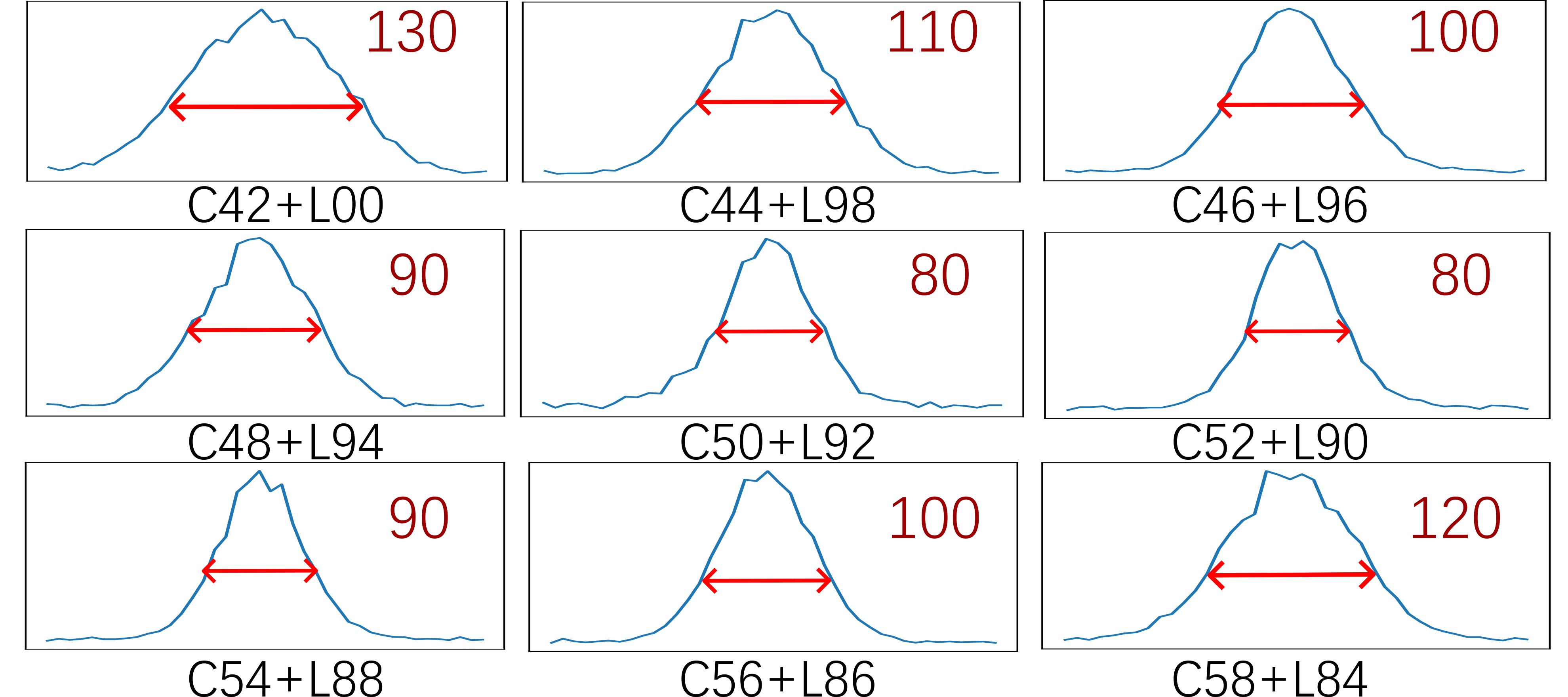}
    \caption{
        \textcolor{black}{Coincidence results after dispersion compensation for 301 km fiber. After dispersion compensation, the time jitter for each channel ranges from 80 ps to 130 ps. In this demonstration, 9 pairs of 200 GHz DWDM are selected, ranging from C42 + L00 to C58 + L84, where the channel names denote the center frequency according to the International Telecommunication Union (ITU) standard.}}
    \label{fig:CD}
\end{figure}

\textcolor{black}{Typically, we analyze the dispersion characteristics of each part, residual dispersion and final timing uncertainty of different DWDM channels based on the results of coincidences for 301 km fiber link, yielding the data in Table \ref{tab:301 km dispersion compensation} and Fig. \ref{fig:CD}. In Table \ref{tab:301 km dispersion compensation}, the second-to-last row represents the timing uncertainty calculated based on the residual dispersion and the baseline 60 ps timing uncertainty, while the last row shows the experimental timing uncertainty results for different channels. The finally timing uncertainty is 80 ps in the optimal central DWDM channel caused by the compensation imperfections and higher-order dispersion.}

\textcolor{black}{Based on this analysis, in future QKD experiments, precise dispersion compensation can be tailored for DWDM channels with different central wavelengths to achieve the optimal SKR.}

\begin{table*}[htbp]
    \hspace*{-3.5cm} 
    \caption{\bf Dispersion compensation analysis for 301 km. Unit: ps.}
    \resizebox{\textwidth}{!}{
    \begin{tabular}{cccccccccc}
    \hline
    \hline
    DWDM channel & C42L00 & C44L98 & C46L96 & C48L94 & C50L92 & C52L90 & C54L88 & C56L86 & C58L84\\
    \hline
    Fiber chromatic dispersion & 6489.9 & 6478.5 & 6467.3 & 6456.1 & 6445.0 & 6433.8 & 6422.9 & 6412.0 & 6401.2\\
    DCF & -221.5 & -210.3 & -209.1 & -207.9 & -206.7 & -205.5 
    & -204.3 & -203.1 & -201.2\\
    DCM & -6369.1 & -6333.4	& -6297.8 & -6262.3	& -6226.8 & -6191.3 & -6156.0 & -6120.7 & -6085.7\\
    Residual chromatic dispersion & -90.7 & -65.2 & -39.7 & -14.1 & 11.5  & 37.0 & 62.6 & 88.2 & 113.6\\
    Timing uncertainty (analysis) & 109 & 89 & 72 & 62 & 61 & 70 & 87 & 107 & 128\\
    Timing uncertainty (experiment) & 130 & 110 & 100 & 90 & 80 & 80 & 90 & 100 & 120\\ 
    \hline
    \hline
    \end{tabular}}
    \label{tab:301 km dispersion compensation}
\end{table*}

\subsection*{S9. Loss analysis}

\textcolor{black}{
We analyzed the losses in the 201 km, 301 km and 404 km experiment in Table~\ref{tab:loss}.
For three different experimental configurations, there are common attenuation factors as well as attenuation differences caused by the varying ultra-low-loss fiber (ULLF), dispersion compensation module (DCM), and dispersion compensation fiber (DCF) configurations. The common attenuation factors include the following:
The combined loss from the entanglement source and superconducting nanowire single-photon detectors (SNSPDs) is around 20 dB, corresponding to a coincidence efficiency of approximately $\sim$ 10\%.
This coincidence efficiency includes the detection efficiency of approximately 70\%, the parametric photon transmission and coupling efficiency of around 22\%, and other optical losses of about 67\%. The polarization analysis module (PAM) contributes an additional 3 dB of loss.
}

\textcolor{black}{Additionally, the length of DCF required to compensate for 10 km of dispersion is 1.6 km, exhibiting an attenuation of 1-2 dB. For the DCM, each module introduces an attenuation of 3–4 dB. In the case of the ULLF, to facilitate switching between different configurations, a 50 km fiber spool was used and connected via a Fiber Optic Adapter (FOA). In the 201 km QKD experiment, the ULLFs with FOAs introduced an attenuation of 32 dB, DCMs contributed 6 dB of attenuation, and the DCF introduced 1 dB of attenuation. In the 301 km QKD experiment, the ULLFs with FOAs introduced 50 dB of attenuation, the DCMs introduced 10 dB of attenuation, and the DCF introduced 1 dB of attenuation. In the 404 km QKD experiment, the ULLFs with FOAs introduced 68 dB of attenuation, the DCMs introduced 14 dB of attenuation, and the DCFs introduced 5 dB of attenuation.}

\textcolor{black}{Finally, the total losses of 201 km, 301 km and 404 km are 62 dB, 84 dB and 110 dB.}

\begin{table*}[htbp]
    \centering
    \caption{\bf Loss analysis for QKD. Unit: dB.}
    \begin{tabular}{cccc}
    \hline
    \hline
    Item & 201 km & 301 km & 404 km \\
    \hline
    Entanglement source & 17 & 17 & 17\\
    Fiber & 32 & 50 & 68\\
    SNSPD & 3 & 3 & 3\\
    PAM & 3 & 3 & 3\\
    Dispersion compensation & 7 & 11 & 19\\
    Total & 62 & 84 & 110\\
    \hline
    \hline
    \end{tabular}
    \label{tab:loss}
\end{table*}

\subsection*{S10. Quantum bit error rate analysis}

\begin{figure}[htbp]
    \centering
    \includegraphics[width=6cm]{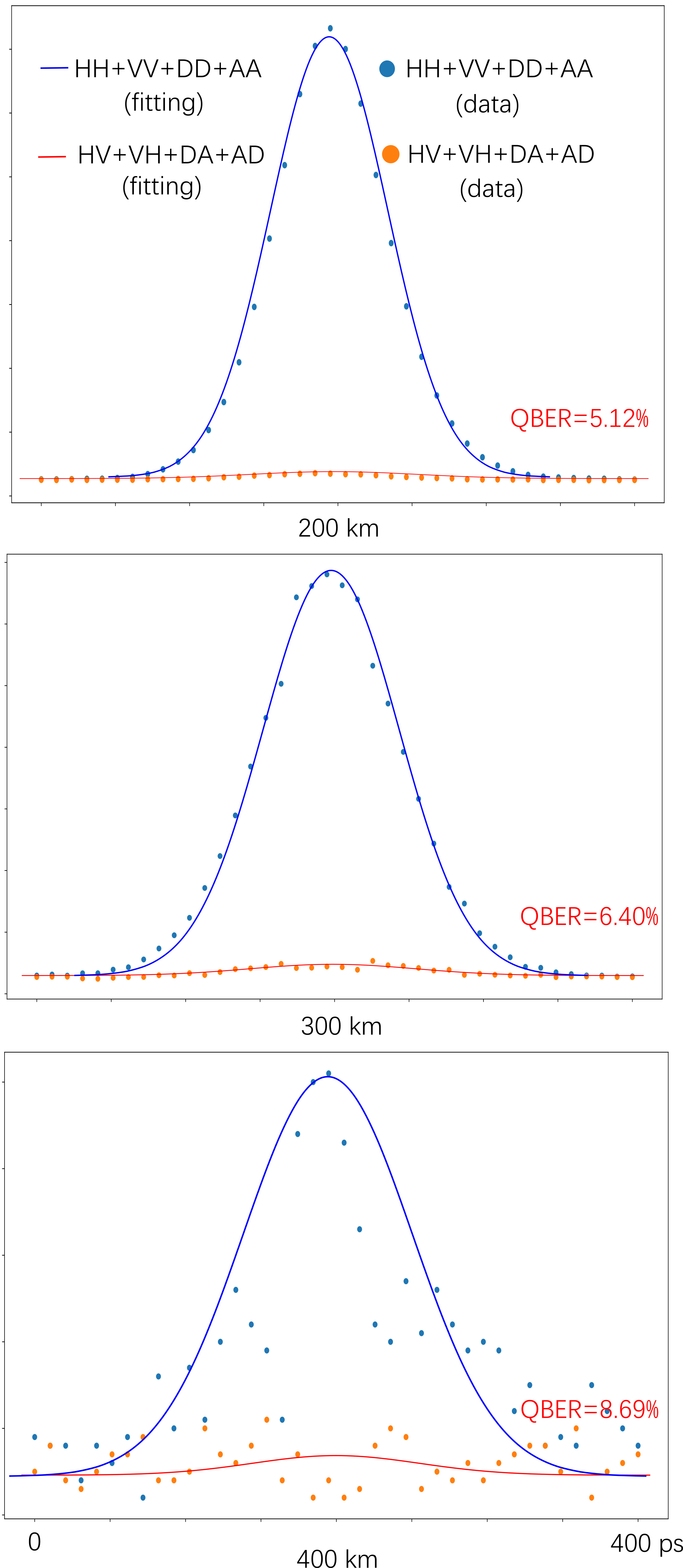}
\caption{
    \textcolor{black}{The QBER result for the summed DWDM channels over 201 km, 301 km, 404 km fiber link. The signal data consists of HH, VV, DD, and AA coincidences, while the error data includes HV, VH, DA, and AD coincidences. The total QBERs for the distances of 201 km, 301 km, and 404 km were measured to be 5.12\%, 6.40\%, and 8.69\%, respectively. The blue and red curves represent the Gaussian fitting results for the signal and error data, respectively.}
}
    \label{fig:error}
\end{figure}

Several factors contribute to the final QBER in the experiment, including imperfect polarization alignment, accidental coincidence counts, and dark counts from the detectors. \textcolor{black}{The final results are depicted in Fig. \ref{fig:error}. The QBERs at 201 km and 301 km were 5.12\% and 6.40\%, respectively, primarily influenced by a combination of imperfect polarization alignment and accidental coincidence counts. For the 404 km link, imperfect polarization alignment and accidental coincidences were responsible for approximately 6.40\% of the QBER. Additionally, at a wavelength of 1540 nm, the signal count rate per channel ranged from 90 to 120 counts/s, including dark counts of approximately 30 counts/s. The dark counts contributed an additional 2.29\% for the total of 8.69\%.}
\textcolor{black}{Our experimental results close to the theoretically predicted QBERs for distances of 201 km, 301 km, and 404 km, which are 5.05\%, 5.83\%, and 8.12\%, respectively.}

\subsection*{S11. Finite key rate calculation}

We used an improved finite-key security analysis to evaluate the SKR. 
We need to find a bounded set of four-dimensional real vector $\vec{x}=(\alpha,\beta,\nu,\xi)$ to solve the goal programming as follows:
\begin{align*}
&\max_{\vec{x}\in \mathbb{R}^{4}}   &&\ell=\left \lfloor \alpha m \right \rfloor \nonumber\\
&\text{such that} &&2^{-t}+2\varepsilon _{\text{pe}}(\nu,\xi)+\varepsilon _{\text{pa}}(\nu)\le \varepsilon _{\text{QKD}},\nonumber\\
& &&\varepsilon _{\text{pa}}(\nu)=\frac{1}{2}\sqrt{2^{-n(1-h_2(\delta+\nu))+r+t+\ell}},\nonumber\\
& &&\varepsilon _{\text{pe}}(\nu,\xi)^2\!=\!\exp(\!-\frac{2mk{\xi}^2}{n+1})\!+\!\exp(\!-2\gamma[n^2{\nu'}^2\!\!-\!1]),\nonumber\\
& &&\varepsilon _{\text{QKD}}\!=\!10^{-s},2^{-t}\!=\!10^{-(s+2)},k\!=\!\left \lfloor \beta m \right \rfloor ,n\!=\!m\!-\!k,\nonumber\\
& &&\gamma\!=\!\max\{\frac{1}{n\!+\!1}\!+\!\frac{1}{k\!+\!1},\frac{1}{m_{\text{err}}\!+\!1}\!+\!\frac{1}{m\!-\!m_{\text{err}}\!+\!1}\} ,\nonumber\\
& &&m_{\text{err}}=\left \lceil m(\delta+\xi) \right \rceil,\nu'=\nu-\xi,n^2{\nu'}^2>1,\nonumber\\
& &&\alpha \in [0,1],\beta\in(0,1/2],0<\nu<\xi<1/2-\delta.
\end{align*}
 The block length $m$ (sifted key length) and tolerated QBER $\delta$ are obtained from experimental data, and we set the security parameter $\varepsilon _{\text{QKD}}=1\times 10^{-9}$. Noticeably, our objective is to find the maximum secret key length $\ell$; thus $\alpha$ here represents the secret key rate. $2^{-t}$, $\varepsilon _{\text{pe}}(\nu,\xi)$ and $\varepsilon _{\text{pa}}(\nu)$ are the error functions due to error correction,  \textcolor{black}{parameter estimation and privacy amplification}, respectively. In addition, $k$ is the number of sifted bits allocated to parameter estimation, and $r=1.09h_2(\delta)$ is the expected error correction leakage, where the binary entropy function $h_2=-x\text{log}x-(1-x)\text{log}(1-x)$.

\subsection*{S12. Asymptotic key rate calculation}

The asymptotic secret key rate $R_Z$ for the post-processed bits in the Z basis is given by:
    \begin{equation}
        R_Z \geq Q_Z[1-f_eH(E_Z)-H(E_X)],
    \end{equation}

where $Q_z$ is the sifted key when Alice and Bob select the Z basis, $f_e$ is the error correction inefficiency. $E_Z$ and $E_X$ are the QBER in the Z and X bases, respectively. The analysis for the X basis is the same. The total asymptotic secret key rate is $R_A= R_Z+ R_X$. An error correction inefficiency factor of $f_e$ = 1.09 was used for the calculations. 


\end{document}